%Paper: cond-mat/9509113
%From: janco@stat.th.u-psud.fr (jancovici)
%Date: Tue, 19 Sep 1995 12:52:43 +0100

\magnification=1200
\headline{\ifnum\pageno=1 \nopagenumbers
\else \hss\number \pageno \fi}

\overfullrule=0pt

\font\boldgreek=cmmib10
\textfont9=\boldgreek
\mathchardef\myrho="091A
\def\bfrho{{\fam=9 \myrho}\fam=1}
\mathchardef\mymu="0916 
\footline={\hfil}
\baselineskip=10pt
\parskip=0.2truecm
\centerline{\bf Universality in Some Classical Coulomb Systems}
\medskip
 \centerline{\bf of Restricted Dimension}\bigskip

\medskip
\centerline{{\bf P. J. Forrester}$^1${\bf ,}\footnote{}{$^1$ Department of
Mathematics,
University of Melbourne, Parkville, Victoria 3052, Australia. E-mail~:
matpjf@maths.mu.oz.au} {\bf B. Jancovici}$^{2}${\bf ,}\footnote{}{$^2$
Laboratoire de Physique
Th\'eorique et Hautes Energies, Universit\'e de Paris XI, b\^atiment 211, 91405
Orsay Cedex, France
(Laboratoire associ\'e au Centre National de la Recherche Scientifique - URA
D0063). E-mail~:
janco, tellez@stat.th.u-psud.fr} {\bf and G. T\'ellez}$^{2}$}

\bigskip
\baselineskip=20pt
\noindent
${\bf Abstract}$ \par
Coulomb systems in which the particles interact through the $d$-dimensional
Coulomb potential
but are confined in a flat manifold of dimension $d - 1$ are considered. The
Coulomb
potential is defined with some boundary condition involving a characteristic
macroscopic
distance $W$ in the direction perpendicular to the manifold~: either it is
periodic of
period $W$ in that direction, or it vanishes on one ideal conductor wall
parallel to the
manifold at a distance $W$ from it, or it vanishes on two parallel walls at a
distance $W$
from each other with the manifold equidistant from them. Under the assumptions
that
classical equilibrium statistical mechanics is applicable and that the system
has the
macroscopic properties of a conductor, it is shown that the suitably smoothed
charge
correlation function is universal, and that the free energy and the grand
potential have
universal dependences on $W$ (universal means independent of the microscopic
detail). The
cases $d = 2$ are discussed in detail, and the generic results are checked on
an exactly
solvable model. The case $d = 3$ of a plane parallel to an ideal conductor is
also
explicitly worked out.

\vbox to 0.1 truecm{}

\vbox to 0.1 truecm{}

\noindent LPTHE Orsay 95-55 \par
\noindent July 1995
\vfill\supereject \noindent {\bf 1. INTRODUCTION} \medskip
The present paper is about Coulomb systems, in which the particles interact
through the
$d$-dimensional Coulomb potential (or some variant of it) but live in a space
of dimension $d
- 1$ (this is what we call restricted dimension). Examples are particles in a
plane
in\-te\-rac\-ting through the usual $1/r$ Coulomb potential, or particles on a
line inte\-racting
through the two-dimensional $-\ln r$ Coulomb potential. We are interested in
the classical
equilibrium statistical mechanics of some of these systems (``classical'' means
that the
quantum effects are disregarded), with the purpose of exhibiting universal
properties
(``universal'' means independent of the microscopic detail). This universality
is closely
related to the universality of the macroscopic electrostatics of conductors.
Therefore,
from the beginning, we assume that the systems under consideration have a
macroscopic
conducting behavior~; this excludes for instance a two-component
one-dimensional log-gas
at a too low temperature, since this gas is then insulating.$^{(1)}$ \par

The charge correlations have already been shown to be universal for several
classes of
conducting systems of restricted dimension. In the simple above-mentioned
cases,$^{(2)}$
the charge-charge correlation function $S(r)$ is
$$S(r) = - {k_BT \over 4 \pi^2 r^3} \eqno(1.1)$$

\noindent in a plane with $1/r$ interactions, and
$$S(r) = - {k_BT \over \pi^2 r^2} \eqno(1.2)$$

\noindent in a log-gas on a line~; $k_B$ is Boltzmann's constant, $T$ the
temperature, and
$r$ the distance. (These expressions (1.1) and (1.2) are only macroscopically
valid~: the
distance $r$ must be large compared to the microscopic scale, and possible
oscillations of
the correlation function have to be smoothed away). Another occurrence of
universality is
for systems with a $d$-dimensional Coulomb interaction confined in some
appropriate
$d$-dimensional domain~; then, not only are there universal surface charge
correlations,$^{(2)}$ but also the free energy and the grand potential have
universal
finite-size corrections.$^{(3-5)}$ Here we shall be concerned with infinite
flat systems
of restricted dimension $d - 1$, with some finite-size effect brought in by a
boundary
condition on the electric potential in the $d$th dimension. Universal
behavior will be found both for the charge correlations and the free energy or
the grand
potential. For $d = 2$, these generic properties will be checked on a solvable
model. \par

For instance, in Section 2, we consider a conducting log-gas on an infinite
straight line
(the $x$-axis) with the logarithmic interaction modified by the constraint
that it be periodic of period $W$ in the transverse direction $y$. Then, for
macroscopic
distances, the smoothed charge-charge correlation function is found to be
$$S(x) = - {k_BT \cosh (\pi x/W) \over W^2 \sinh^2 (\pi x/W)} \eqno(1.3)$$

\noindent while the free energy per unit legnth $f$ and the grand potential per
unit
length $\omega$ exhibit a finite-$W$ correction as $W \to \infty$~:
$$\eqalignno{
&f(W) = f(\infty ) + k_BT {\pi \over 8W} + o(W^{-1}) &(1.4a) \cr
&\omega (W) = \omega (\infty ) + k_BT {\pi \over 8W} + o(W^{-1}) &(1.4b) \cr
}$$

\noindent These generic results (1.3) and (1.4) are checked in the special case
of a
one-component log-gas which is a (partially) exactly solvable model. \par

Similar results are obtained for a conducting line at a distance $W$ from an
ideal conductor
in Section 3, and for a conducting line with ideal conductors on each side of
it in Section
4. An example of higher dimension is considered in Section 5~: a conducting
plane at a
distance $W$ from an ideal conductor.

 \vfill \supereject
\noindent {\bf 2. LOG-GAS ON A LINE WITH TRANSVERSE PERIODIC BOUNDARY
CONDITION} \medskip
{\bf 2.1. \underbar{The system}} \medskip
The two-dimensional Coulomb interaction between two point charges $q$ and $q'$
located in the
$xy$ plane at ${\bf r} = (x, y)$ and ${\bf r}' = (x', y')$ is $qq' G_0({\bf r},
{\bf r}')$,
with $G_0({\bf r}, {\bf r}')$ a solution of the Poisson equation
$$\Delta G_0({\bf r}, {\bf r}') = - 2 \pi \delta ({\bf r} - {\bf r}')
\eqno(2.1)$$

\noindent In the strip $- W/2 \leq y, y' \leq W/2$, a solution of (2.1) with
periodic
boundary conditions at $y = - W/2$ and $y = W/2$ is
$$G_0({\bf r} , {\bf r}') = - \ln \left | {W \over \pi} \sinh {\pi (z - z')
\over W}
\right | \eqno(2.2)$$

\noindent where $z = x + iy$ and $z' = x' + iy'$ are the complex coordinates~;
the factor
$W/\pi$ in (2.2) ensures that, in the limit $W \to \infty$, one recovers the
usual
two-dimensional Coulomb interaction $- \ln |z - z'|$. We consider some
one-dimensional
system of charges, on the $x$ axis, with the interaction (2.2), where now $z =
x$ and $z' =
x'$~; some short-range interaction might also be present. It may be noted that
the interaction
$$v(x - x') = - \ln \left | {W \over \pi} \sinh {\pi \over W} (x - x') \right |
\eqno(2.3)$$

\noindent on the $x$ axis interpolates between a logarithmic interaction $- \ln
|x -
x'|$ at short distances and a linear interaction (one-dimensional Coulomb
interaction) $-
( \pi /W)|x - x'|$ at large distances. \par

The system is assumed to have the properties which characterize a conductor,
and to be
globally neutral. The period $W$ is macroscopic. \vfill \supereject

{\bf 2.2. \underbar{Electric potential correlations}} \medskip
For investigating the charge correlations and the thermodynamics, we first need
information
about the correlations of the electric potential. Let $\phi ({\bf r})$ be the
microscopic
electric potential created at ${\bf r}$ by the charges of the system and let us
consider the
correlation function $<\phi ({\bf r}) \phi ({\bf r}')>$. For ${\bf r}$ and
${\bf r}'$ at
macroscopic distances from the $x$ axis, this correlation function can be
obtained by the
method described in ref. 1, using linear response theory and the conducting
behavior
assumption, as follows. Let us put an infinitesimal test charge $q$ at ${\bf
r}'$. Its
interaction with the system is described by a Hamiltonian $q \phi ({\bf r}')$,
and by linear
response theory the average potential at ${\bf r}$ created by the charges of
the system is
changed by
$$\delta \phi ({\bf r}) = - \beta q <\phi ({\bf r}) \phi ({\bf r}')>^T
\eqno(2.4)$$

\noindent where $\beta = 1/k_BT$ and $<\cdots>^T$ means a truncated statistical
average
($<AB>^T\break \noindent = <AB> - <A><B>$)~; here however $<\phi ({\bf r})> =
0$, and the mark
``truncated'' is superfluous. The total potential change at ${\bf r}$ is
$qG({\bf r}, {\bf
r}')$, where $G({\bf r}, {\bf r}')$ is given by the macroscopic electrostatics
of conductors,
i.e. $G$ is the solution of the Poisson equation for a point charge $q$ at
${\bf r}'$, with the
constraint that it is periodic of period $W$ in $y$ and that it vanishes on the
$x$ axis $y =
0$~:
$$\eqalignno{
&G({\bf r}, {\bf r}') = - \ln \left | {\sinh {\pi \over 2W} (z - z') \over
\sinh {\pi
\over 2W} (z - \bar{z}')} \right | \qquad \hbox{if $yy' > 0$} &(2.5a) \cr
&G({\bf r} , {\bf r}') = - \ln \left | {\cosh {\pi \over 2W} (z - z') \over
\cosh {\pi
\over 2W} (z - \bar{z}')} \right | \qquad \hbox{if $yy' < 0$} &(2.5b) \cr
}$$

\noindent ($G$ does not vanish for $yy' < 0$, i.e. the conducting line does
{\it not}
screen the regions $y > 0$ and $y < 0$ from each other~; this is an effect of
the periodic
boundary condition which connects these regions across the boundaries at $y =
\pm W/2$). That
part of the total potential change which is created by the charges of the
system is
$$\delta \phi ({\bf r}) = q \left [ G({\bf r}, {\bf r}') - G_0({\bf r}, {\bf
r}') \right ]
\eqno(2.6)$$

\noindent From (2.4) and (2.6), one obtains for the correlation function
$$\beta <\phi ({\bf r}) \phi ({\bf r}')>^T = q \left [ G_0({\bf r}, {\bf r}') -
G({\bf r}, {\bf
r}') \right ] \eqno(2.7)$$

\noindent in terms of (2.2) and (2.5). \par \medskip

{\bf 2.3. \underbar{Charge correlations}} \medskip
Let $\sigma (x)$ be the charge per unit length on the $x$ axis and let the
ma\-cros\-co\-pi\-cally smoothed charge-charge correlation function be $S(x -
x') = <\sigma
(x) \ \sigma (x')>^T$. Since $2 \pi \sigma (x)$ is equal to the discontinuity
of the electric
field component $-\partial \phi /\partial y$ across the $x$ axis, and since
$G_0$ does not
contribute to that discontinuity, one finds from (2.7)
$$\beta \ S(x - x') = 2 \lim_{y,y' \to 0} \left [  - \left . {\partial^2G({\bf
r}, {\bf r}')
\over \partial y \ \partial y'} \right |_{yy'>0} + \left . {\partial^2 G({\bf
r}, {\bf r}')
\over \partial y \ \partial y'} \right |_{yy'< 0} \right ] \eqno(2.8)$$

\noindent Using (2.5) in (2.8) gives (1.3). \par

An alternative derivation of (1.3) can be given by considering only the
$x$-axis system, with
the interaction (2.3) given, without any reference to the ``outside world''.
One starts with
the assumption that an external infinitesimal linear charge density $q \exp
(ikx)$ is
perfectly screened (for a wave-number $k$ small enough, i.e. macroscopic).
Therefore the
system responds by creating a charge density $- q \exp (ikx)$. The interaction
Hamiltonian of
the external charge with the system is $q \tilde{v}(k) \ \tilde{\sigma}(-k)$,
where
$\tilde{v}(k)$ and $\tilde{\sigma}(-k)$ are the Fourier transforms of $v(x)$
and $\sigma
(x)$ respectively. By linear response theory, $- q = - \beta q \tilde{v}(k)
<\tilde{\sigma}(-k) \tilde{\sigma}(k)>^T$, i.e.

$$\beta \tilde{S}(k) = {1 \over \tilde{v}(k)} \eqno(2.9)$$

\noindent where $\tilde{S}(k) = < \tilde{\sigma}(-k) \tilde{\sigma}(k)>^T$ is
the Fourier
transform of $S(x)$. Our definition of the Fourier transforms is, for instance,
$$\tilde{v}(k) = \int_{- \infty}^{\infty} dx \ e^{-ikx} \ v(x) \eqno(2.10)$$

\noindent Using (2.3) in (2.10) gives (in the sense of distributions)
$$\tilde{v}(k) ={\pi \over k} {\rm ctnh} \ {Wk \over 2} \eqno(2.11)$$

\noindent and, from (2.9), (1.3) follows. \par \medskip

{\bf 2.4. \underbar{Free energy or grand potential}} \medskip
The thermodynamic potential to be considered is the free energy per unit length
$f$ if the
canonical ensemble is used, or the grand potential per unit length $\omega$ if
the grand
canonical ensemble is used. As a starting point we consider the derivative
$\partial
f/\partial W$ (at constant densities) or $\partial \omega/\partial W$ (at
constant
fugacities). \par

Let us draw some line parallel to the $x$ axis~; this line divides the plane
into two regions
$R$ on its right and $L$ on its left. The derivative $\partial f/\partial W$ or
$\partial
\omega/\partial W$ is the force per unit length that region $R$ exerts on
region $L$, i.e.
the $T_{yy}$ component of the Maxwell stress tensor~:
$${\partial f \over \partial W} = {\partial \omega \over \partial W} = T_{yy'}
= {1 \over 4
\pi} <E_y({\bf r})^2 - E_x({\bf r})^2> \eqno(2.12)$$

\noindent where ${\bf E}({\bf r}) = - \nabla \phi ({\bf r})$ is the electric
field at ${\bf
r}$ ($T_{yy}$ should be independent of ${\bf r}$). Since $<{\bf E}({\bf r})> =
0$, we can
replace $<\cdots >$ by $<\cdots >^T$ in (2.12), and using the derivative of
(2.7) with respect
to ${\bf r}$ and ${\bf r}'$ (with the limit ${\bf r}' = {\bf r}$ taken at the
end of the
calculation) one finds
$$\beta \ T_{yy} = {- \pi \over 8W^2} \eqno (2.13)$$

\noindent By integration of (2.12), one obtains (1.4). \par

Alternatively, (1.4) can be derived by considering only the $x$-axis system
with the
interaction $v(x)$. The partition or grand partition function depends on $W$
through $v(x)$,
and deriving it with respect to $W$ gives a statistical average of $\partial
v(x)/\partial
W$ to be taken with the charge correlation function $S(x)$ (since $\partial
v(x)/ \partial
W$ vanishes at $x = 0$, the self-part of $S(x)$ may or may not be kept)~:
$$\beta {\partial f \over \partial W} = \beta {\partial \omega \over \partial
W} = {1 \over
2} \int_{- \infty}^{\infty} S(x) {\partial v(x) \over \partial W} dx = {1 \over
2} \int_{-
\infty}^{\infty} \tilde{S}(k) {\partial \tilde{v}(k) \over \partial W} {dk
\over 2 \pi}
\eqno(2.14)$$

\noindent Using (1.3) and (2.3), or (2.9) and (2.11), in (2.14), one recovers
the result
$\beta \partial f/\partial W = \beta \partial \omega/\partial W = - \pi /8 W^2$
and (1.4). \par
\medskip

{\bf 2.5. \underbar{Solvable model}} \medskip
The one-dimensional one-component plasma with the interaction (2.3) is a
solvable model for
special value(s) of the temperature. The model consists of particles of charge
$e$, with a
particle density $\eta$, and a uniform background of charge density $-e \eta$
which ensures
overall neutrality. \par

The correlations are known at the special temperature such that $\Gamma :=
\beta e^2 = 2$~;
they are the same ones as for some quantum zero-temperature system.$^{(6)}$
{}From eqs.
(3.13a) and (3.21a)$^3$\footnote{}{$^3$ A factor 2 is missing in the
denominator of eq.
(3.21a) in ref. 6.} of ref. 6,
$$\beta \ S(x, x') = - 2 \left [ {1 \over W} {\widehat{\ell}(x) \
\widehat{\ell}(-x') \ -
\widehat{\ell}(-x) \ \widehat{\ell}(x') \over \theta '_1 \left ( 0;e^{-2 \pi
\eta W}
\right ) 2 \sinh  ( \pi |x - x'|/W) } \right ]^2 + 2 \eta \ \delta (x - x')
\eqno(2.15)$$

\noindent where
$$\widehat{\ell}(x) = e^{-\pi x/2W} \theta_1 \left ( {\pi \over 4} + \pi \eta x
; e^{-\pi
\eta W} \right ) \eqno(2.16)$$

\noindent ($\theta_1$ is a Jacobian theta function). The charge-charge
correlation function
$S(x, x')$ does not depend only on the distance $|x - x'|$ because the system
has a
crystalline structure. For $q$ small,
$$\theta_1(u;q) \sim 2q^{1/4} \sin u \eqno(2.17)$$

\noindent For $W$ large, (2.17) can be used in (2.15) and (2.16), giving
\vfill \supereject
$$\beta \ S(x, x') \sim - {1 \over W^2 \sinh^2 {\pi (x - x') \over W}} \left [
2 \sin^2
\left ( {\pi \over 4} + \pi \eta x \right ) \sin^2 \left ( {\pi \over 4} - \pi
\eta x'
\right ) e^{- \pi(x - x')/W} \right .$$
$$\left . + 2 \sin^2 \left ( {\pi \over 4} - \pi \eta x \right ) \sin^2 \left (
{\pi \over 4} +
\pi \eta x' \right ) e^{\pi (x - x')/W} - \cos 2 \pi \eta x \cos 2 \pi \eta x'
\right ] \ \ , x
\not= x'  \eqno(2.18)$$

\noindent The macroscopically smoothed $S(x - x')$ is obtained by averaging out
the
microscopic oscillations in (2.18), i.e. replacing the $\sin^2$ terms by 1/2
and the $\cos$
terms by 0, which gives
$$\beta \ S(x - x') \sim - {\cosh [\pi (x - x')/W] \over W^2 \sinh^2 [\pi (x -
x')/W]}
\eqno(2.19)$$

\noindent in agreement with (1.3). \par

The free energy can be computed at the special temperatures such that $\Gamma$
= 1, 2, 4,
starting with the known partition function$^{(7)}$ for a finite system of $N$
particles
on a line of length $L$ along the $x$ axis, and an interaction which is
periodic of
period $W$ in the $y$ direction and also periodic of period $L$ in the $x$
direction~; on
the line, this interaction is
$$v(x - x') = - \ln \left | \theta_1 \left ( {\pi (x - x') \over L} ; e^{- \pi
W/L}
\right ) \right | \hbox{+ constant} \eqno(2.20)$$

\noindent (with a suitable choice of the constant, (2.20) goes to (2.3) as $L
\to \infty$).
Including in the energy the particle-particle, particle-background and
background-background
interactions, one finds for the partition function
$$Z_N (\Gamma ) = \left [ {\pi \theta '_1 (0;q) \over L} \right ]^{N\Gamma /2}
\ q^{-N^2\Gamma
/8} \left [ \prod_{n=1}^{\infty} \left ( 1 - q^{2n} \right ) \right
]^{-N^2\Gamma /2} {1 \over
N!} C_{N \Gamma} \eqno(2.21)$$

\noindent where $q := \exp (- \pi W/L)$ and $C_{N \Gamma}$ is the configuration
integral
$$C_{N \Gamma} := \prod_{\ell = 1}^N \int_0^L dx_{\ell} \prod_{1 \leq j < k
\leq N} \left
| \theta_1 \left ( {\pi (x_k - x_j) \over L} ; q \right ) \right |^{\Gamma}
\eqno(2.22)$$

\noindent $C_{N \Gamma}$ has been computed in ref. 7 for $\Gamma$ = 1, 2, 4. In
the
thermodynamic limit $N \to \infty$, $L \to \infty$, $N/L = \eta$ fixed, one
finds for the
free energy per unit length $f(\Gamma ; W) = - \lim L^{-1} \ln Z_N(\Gamma )$
$$\beta  f(1;W) = {1 \over 2} \eta \ln {\eta \over 2} + {\pi \over 12W} - \eta
\int_0^{1/2} dt \ln \left [ \sum_{n = - \infty}^{\infty} {e^{-2 \pi W \eta (n^2
+ 2nt)}
\over n + t} \right ] \eqno(2.23a)$$

$$\beta f(2;W) = - \eta \ln (2 \pi ) + {\pi \over 8W} - \eta
\sum_{n=1}^{\infty} \ln \left ( 1
- e^{-4\pi W \eta n} \right ) \eqno(2.23b)$$

$$\beta  f(4;W) = - \eta \ln (8\pi^2 \eta ) + {\pi \over 12W} - \eta \int_0^1
dt \ln
\left [ \sum_{n=-\infty}^{\infty} (2n + t) e^{-4 \pi W \eta (n^2 + nt)} \right
]
\eqno(2.23c)$$

\noindent In the large $-W$ limit, in all cases,
$$\beta f(\Gamma; W) \sim \beta f (\Gamma ; \infty ) + {\pi \over 8W}
\eqno(2.24)$$

\noindent
$$\beta f(\Gamma ; \infty ) = \eta \left [ \left ( 1 - {\Gamma \over 2} \right
) \ln (2
\pi \eta ) - {\Gamma \over 2} \ln {\Gamma \over 2} + {\Gamma \over 2} + \ln
\left ( {\Gamma
\over 2} \right ) ! - \ln (2 \pi ) - 1 \right ] \eqno(2.25)$$
\par \medskip

\noindent {\bf 3. LOG-GAS ON A LINE PARALLEL TO AN IDEAL CONDUCTOR} \medskip
{\bf 3.1. \underbar{The system}} \medskip
The region of interest is the half-plane $y > 0$. The electric potential is
constrained to
vanish on the $x$ axis, i.e. the $x$ axis is an ideal conductor at zero
potential. A
solution of (2.1) with that boundary condition is
$$G_0({\bf r}, {\bf r}') = - \ln \left [ {z - z' \over z - \bar{z}'} \right |
\eqno(3.1)$$

\noindent We consider some one-dimensional system of charges, on the line $y =
W$, with the
cor\-res\-ponding interaction
$$v(x - x') = - {1 \over 2} \ln {(x - x')^2 \over (x - x')^2 + 4W^2}
\eqno(3.2)$$

\noindent plus perhaps some short-range interaction. It may be noted that this
interaction
interpolates between $- \ln |x - x'|$ at short distances and $2W^2/(x - x')^2$
at large
distances. $W$ is a macroscopic distance, and the system is assumed to have the
properties of
a conductor. \par \medskip

{\bf 3.2. \underbar{Correlations}} \medskip
We follow the same steps as in Section 2. Let us first consider the case when
the conducting
system on the line $y = W$ is kept at zero macroscopic potential. Now $G$ is
the solution of
the Poisson equation for a point charge $q$ at ${\bf r}'$ with the constraint
that it
vanishes on the lines $y = 0$ and $y = W$~:
$$G({\bf r}, {\bf r}') = - \ln \left | {\sinh {\pi \over 2W} (z - z') \over
\sinh {\pi \over
2W} (z - \bar{z}')} \right | \qquad \hbox{if $0 < y, y' < W$} \eqno(3.3a)$$

$$G({\bf r}, {\bf r}') = - \ln \left | {z - z' \over z - \bar{z}'- 2iW} \right
| \qquad
\hbox{if $y, y' > W$} \eqno(3.3b)$$

$$G({\bf r}, {\bf r}') = 0 \qquad
\hbox{if $0 < y  < W, y' > W$ \ ; \ or $y > W$, $0 < y' <W$} \eqno(3.3c)$$

\noindent From the analog of (2.8) one now obtains the universal correlation
function
$$\beta \ S(x) = - {1 \over 2 \pi^2 x^2} - {1 \over 8W^2 \sinh^2 {\pi x \over
2W}}
\eqno(3.4)$$

An alternative direct derivation of (3.4) uses (2.9) and the Fourier transform
$$\tilde{v}(k) = {\pi \over |k|} \left ( 1 - e^{-2W|k|} \right ) \eqno(3.5)$$
\par \medskip

{\bf 3.3. \underbar{Free energy or grand potential}} \medskip
Still assuming that the line $y = W$ is kept at zero macroscopic potential, we
can use (2.7)
(with $< \cdots >^T = < \cdots>$), (3.1), and (3.3a) for computing the $T_{yy}$
component of
the stress tensor at some point ${\bf r}$ between the ideal conductor and the
conducting
line. One finds
$$\beta \ T_{yy} := {\beta \over 4 \pi} <E_y({\bf r})^2 - E_x ({\bf r})^2> =
{\pi \over
24W^2} \eqno(3.6)$$

\noindent The $W$-dependence of the free energy or grand potential is given by
$${\partial f \over \partial W} = {\partial \omega \over \partial W} = T_{yy}
\eqno (3.7)$$

\noindent provided $f$ and $\omega$ are properly defined with a Hamiltonian
which includes
the self-energy interaction $(1/2) q^2 \ln (2W)$ of each particle of charge $q$
with its
image. It should be noted that, in the limit $W \to
\infty$, this self-energy and the two-body interaction (3.2) generate a
well-defined
total Hamiltonian containing only a two-body interaction $- \ln |x - x'|$~;
therefore $f$
and $\omega$ are expected to have well-defined limits as $W \to \infty$. From
(3.6) and (3.7),
one finds the large-$W$ expansions
$$\beta f(W) = \beta f(\infty ) - {\pi \over 24W} + o \left ( W^{-1} \right )
\eqno(3.8a)$$

$$\beta \omega (W) = \beta \omega (\infty ) - {\pi \over 24W} + o \left (
W^{-1} \right )
\eqno(3.8b)$$

\noindent with the universal finite-$W$ correction $- \pi /24W$. \par

Alternatively, (3.8) can be obtained from (2.14) by using either (3.2) and
(3.4) (with the
integral on $x$ in (3.8) defined as its finite part), or (2.9) and (3.5). Since
one must keep
the self-contribution from the $(1/2) \ln [(x - x')^2 + 4W^2]$ part of (3.2),
it is indeed
appropriate to use in (2.14) the full correlation function $S(x)$ or
$\tilde{S}(k)$ which
includes the self part~; when understood as the Fourier transform of $\beta
\tilde{S}(k) =
1/\tilde{v}(k)$ in the sense of distributions, (3.4) does represent the full
$\beta S(x)$.
\vfill \supereject

{\bf 3.4. \underbar{Non-zero potential difference}} \medskip
We now consider the more general case when the system on the line $y = W$ is
kept at a
non-zero macroscopic potential $\Phi$~; correspondingly, there is on that line
an average
linear charge density
$$\sigma = {\Phi \over 2 \pi W} \eqno(3.9)$$
\noindent (3.9) is the equivalent of the familiar charge-potential relation in
a plane
condenser. \par

Provided one defines the charge correlation function as $S(x) = <\sigma (0)
\sigma (x)>^T$,
where the $T$ (truncated) sign is now relevant, the calculation of $S(x)$ is
unchanged and
(3.4) is still valid. \par

The free energy and grand potential however get additional terms. $<E_y({\bf
r})>$ is no
longer zero, and (3.6) must be replaced by
$$\eqalignno{
\beta T_{yy} &:= {\beta \over 4 \pi} <E_y({\bf r})^2 - E_x ({\bf r})^2> =
{\beta \over 4 \pi}
\left [ <E_y({\bf r})^2 - E_x({\bf r})^2>^T + <{\bf E}_y({\bf r})> \right ] \cr
&= {\pi \over 24W^2} + {\beta \Phi^2 \over 4 \pi W^2} = {\pi \over 24W^2} + \pi
\beta
\sigma^2 &(3.10) \cr
}$$

\noindent From $(\partial f/\partial W)_{\sigma} = T_{yy}$ one obtains
$$\beta f(W) \sim \beta f(\infty ) - {\pi \over 24W} + \beta \pi \sigma^2 W =
\beta f(\infty
) - {\pi \over 24W} + {\beta \Phi^2 \over 4 \pi W} \eqno(3.11a)$$

\noindent while from $(\partial \omega /\partial W)_{\Phi} = T_{yy}$ one
obtains
$$\beta \omega (W) \sim \beta \omega (\infty ) - {\pi \over 24W} - {\beta
\Phi^2 \over 4 \pi
W} \eqno(3.11b)$$

\noindent It is well known that the macroscopic electrostatic energy $\Phi^2/4
\pi W = \pi
\sigma^2$ must come with different signs in $f$ and in $\omega$. \par \medskip

{\bf 3.5. \underbar{Solvable model}} \medskip
The one-dimensional one-component plasma on a line parallel to an ideal
conductor is a
solvable model,$^{(8,9)}$ at the special temperature $\Gamma := \beta e^2 =
2$~; it is a
special case of the more general models of two-dimensional plasmas with an
ideal conductor
wall$^{(10)}$ or between two ideal conductor walls.$^{(5)}$ One uses the grand
canonical
ensemble, with a fixed linear charge density $- e \eta$ for the background and
a fugacity
$\zeta$ which governs the particle density. The distance $W$ between the system
and the
ideal conductor can have any value (not necessarily large) to start with. \par

By a simple adaptation of the formalism in previous work,$^{(5, 8, 9)}$ we
obtain for the
grand potential per unit length $\omega$ (including the background self-energy)
$$\beta \omega = - {\rm Tr}\  \ln (1 + K) + 2 \pi \eta^2W \eqno(3.12)$$

\noindent where $K$ is the continuous matrix
$$K(x, x') = i \zeta {e^{4 \pi \eta W} \over x - x' + 2iW} \eqno(3.13)$$

\noindent Using the Fourier transform which diagonalizes $K$
$$\tilde{K}(k) = \int_{- \infty}^{\infty} dx' \ e^{ik(x' - x)} K(x , x') =
\cases {2 \pi
\zeta e^{4 \pi \eta W - 2kW} \  &{\rm if} \ $k > 0$ \cr
\cr
0 &{\rm if} \ $k < 0$ \cr
} \eqno(3.14)$$

\noindent and writing the trace (per unit length) as $(2 \pi)^{-1} \int dk$,
one finds
$$\beta \omega = - \int_0^{\infty} {dk \over 2 \pi} \ln \left [ 1 + 2 \pi \zeta
e^{4 \pi \eta
W - 2kW} \right ] + 2 \pi \eta^2 W \eqno(3.15)$$

For obtaining a large-$W$ expansion ($\eta W \gg 1$) of (3.15), we make the
change of
variable $4 \pi \eta W - 2 kW = - u$, split the integral on $u$ into two
integrals in the
$u$-ranges $(-4 \pi \eta W, 0)$ and $(0, \infty )$, and write
$$2 \pi \eta^2 W =  - {1 \over 4 \pi W} \int_{-4 \pi \eta W}^0 du \ln e^u
\eqno(3.16)$$

\noindent This leads to the still exact expression
$$\beta \omega = - \eta \ln (2 \pi \zeta) - {1 \over 4 \pi W} \left [ \int_{- 4
\pi \eta W}^0
du \ln \left ( 1 + {1 \over 2 \pi \zeta} e^u \right ) + \int_0^{\infty} du \ln
\left ( 1 + 2
\pi \zeta e^{-u} \right ) \right ] \eqno(3.17)$$

\noindent Finally, when $\eta W \gg 1$, we can replace the lowest bound of the
first integral
in (3.17) by $- \infty$ and change $u$ into $- u$, which gives, up to
exponentially small
terms,
$$\eqalignno{
\beta \omega &\sim - \eta \ln (2 \pi \zeta ) - {1 \over 4 \pi W}
\int_0^{\infty} du \ln \left
[ \left ( 1 + {1 \over 2 \pi \zeta } e^{-u} \right ) \left ( 1 + 2 \pi \zeta
e^{-u} \right )
\right ] \cr
&= - \eta \ln (2 \pi \zeta ) - {\pi \over 24W} - {1 \over 8 \pi W} \left [ \ln
(2 \pi \zeta )
\right ]^2 &(3.18) \cr
 }$$

\noindent The corresponding particle density is
$$n = - \zeta {\partial \over \partial \zeta} (\beta \omega ) \sim \eta + {1
\over 4 \pi W}
\ln (2 \pi \zeta ) \eqno(3.19)$$

\noindent and the charge density of the system is $e(n - \eta ) = (e/4 \pi W)
\ln (2 \pi \zeta
)$. The system is neutral when $\zeta = 1/2 \pi$. Otherwise, its average
electric potential
is $\Phi = e(n - \eta )W = (e/4 \pi ) \ln (2 \pi \zeta )$ and therefore, since
$\beta e^2 = 2$,
$$\zeta = {1 \over 2 \pi} e^{\beta e \Phi} \eqno(3.20)$$

\noindent ($\Phi$ contributes a term $e\Phi$ to the chemical potential, as
expected). Using
(3.20) in (3.18), with $\beta e^2 = 2$, shows that (3.11b) is verified in the
present solvable
model. (3.9) is also verified. Furthermore, for $W \to \infty$, $\beta \omega
(\infty ) = -
\eta \ln (2 \pi \zeta )$, $n = \eta$ (independent of $\zeta$), and the
thermodynamic relation
$\beta f = \beta \omega + (\ln \zeta )n$ becomes $\beta f(\infty ) = - \eta \ln
(2 \pi )$, in
agreement with (2.25). \par

The formalism of previous work$^{(5, 8, 9)}$ also gives the correlation
functions in terms of
the continuous matrix $g(x, x')$ defined in matrix notation as
$$g = {K \over 1 + K} \eqno(3.21)$$

\noindent The particle density is
$$n = g(0, 0) \eqno(3.22)$$

\noindent and the charge correlation function is
$$S(x) = - e^2|g(x, 0)|^2 + e^2 n \delta (x) \eqno(3.23)$$

\noindent From (3.21) and (3.14), one finds
$$g(x, 0) = \int_0^{\infty } {dk \over 2 \pi } e^{ikx} {\tilde{K}(k) \over 1 +
\tilde{K}(k)}
= \int_0^{\infty} {dk \over 2 \pi} {e^{ikx} \over 1 + {1 \over 2 \pi \zeta} \
e^{2W(k-2 \pi
\eta )}} \eqno(3.24)$$

\noindent In particular,
$$n = g(0, 0) = {1 \over 4 \pi W} \ln \left ( 1 + 2 \pi \zeta \ e^{4 \pi \eta
W} \right )
\eqno(3.25)$$

In the case of interest $\eta W \gg 1$, (3.25) takes the form (3.19) and (3.24)
can be
rewritten as
$$\eqalignno{
g(x, 0) &\sim \int_0^{\infty} {dk \over 2 \pi} {e^{ikx} \over 1 + e^{2W(k - 2
\pi n)}} \cr
&= - {1 \over 2 \pi ix} {1 \over 1 + e^{-4 \pi nW}} + \int_0^{\infty} {dk \over
2 \pi}
{e^{ikx} \over ix} {W \over 2 \cosh^2 W(k - 2 \pi n )} &(3.26) \cr
}$$

\noindent where an integration by parts has been performed. For $nW \gg 1$, up
to
exponentially small terms, by extending the last integral in (3.26) to the
range $(- \infty ,
\infty )$, one finds
$$g(x, 0) \sim {1 \over 2 \pi i} \left [ - {1 \over x} + {e^{2 \pi inx} \over
{2W \over \pi}
\sinh {\pi x \over 2W}} \right ] \eqno(3.27)$$

\noindent The microscopic correlation function $S(x)$ is obtained by using
(3.19) and (3.27)
in (3.23). It has oscillations of period $n^{-1}$, the average interparticle
distance. When
these oscillations are averaged out, (3.23) with $\beta e^2 = 2$ agrees with
the universal
form (3.4).

\par \medskip
\noindent {\bf 4. LOG-GAS ON A LINE BETWEEN TWO IDEAL CONDUCTORS} \medskip

{\bf 4.1. \underbar{The system}} \medskip
The region of interest is the plane strip $0 \leq y \leq W$. The electric
potential is
cons\-trained to vanish on the lines $y = 0$ and $y = W$, i.e. these lines are
ideal
conductors at zero potential. A solution of (2.1) with that boundary condition
is

$$G_0({\bf r}, {\bf r}') = - \ln \left | {\sinh {\pi \over 2W} (z - z') \over
\sinh {\pi
\over 2W} (z - \bar{z}')} \right | \eqno(4.1)$$

\noindent We consider some one-dimensional system of charges, on the line $y =
W/2$, with
the corresponding interaction
$$v(x - x') = - \ln \left | \tanh {\pi \over 2W} (x - x') \right | \eqno(4.2)$$

\noindent plus perhaps some short-range interaction. Now $v$ interpolates
between $- \ln |x
- x'|$ and $2 \exp [- (\pi /W)|x - x'|]$. Again $W$ is macroscopic and the
system is assumed
to be a conductor~; it is kept at some potential $\Phi$. \par \medskip

{\bf 4.2. \underbar{Correlations}} \medskip
Now, using
$$G({\bf r}, {\bf r}') = - \ln \left | {\sinh {\pi \over 4W} (z - z') \over
\sinh {\pi \over
4W} (z - \bar{z}')} \right | \quad {\rm if} \ 0 < y, y' < {W \over 2}
\eqno(4.3a)$$

$$G({\bf r}, {\bf r}') = 0 \quad {\rm if} \ 0 < y < {W \over 2}, \ {W \over 2}
< y' < W \ ;
\ {\rm or} \ {W \over 2} < y < W, \ 0 < y' < {W \over 2} \eqno(4.3b)$$

\noindent and using the analog of (2.8), one finds the universal correlation
function
$$\beta S(x) = - {1 \over W^2 \sinh^2 {\pi x \over W}} \eqno(4.4)$$

Alternatively, one can use (2.9) and the Fourier transform
$$\tilde{v}(k) = {\pi \over k} \tanh {Wk \over 2} \eqno(4.5)$$
\par \medskip

{\bf 4.3. \underbar{Free energy or grand potential}} \medskip
Now the stress tensor component $T_{yy}$ is found to be such that
$$\beta T_{yy} = {\pi \over 8W^2} + {\beta \Phi^2 \over \pi W^2} \eqno(4.6)$$

\noindent The charge density is
$$\sigma = {2 \Phi \over \pi W} \eqno(4.7)$$

\noindent Thus,
$$\beta f \sim \beta f(\infty ) - {\pi \over 8W} + {\beta \pi \sigma^2 W \over
4} = \beta
f(\infty ) - {\pi \over 8W} + {\beta \Phi^2 \over \pi W} \eqno(4.8a)$$
\noindent and
$$\beta \omega \sim \beta \omega (\infty ) - {\pi \over 8W} - {\beta \Phi^2
\over \pi W}
\eqno(4.8b)$$
\par \medskip

{\bf 4.4. \underbar{Solvable model}} \medskip
The one-component plasma is again a solvable model, with the present boundary
conditions,
when $\Gamma := \beta e^2 = 2$. The formalism of ref. 5 is applicable, with now
$$K(x, x') = {\pi \zeta \over 2W} {e^{\pi \eta W} \over \cosh \left [ {\pi
\over 2W} (x - x')
\right ]} \eqno(4.9)$$

\noindent and
$$\tilde{K}(k) = \pi \zeta {e^{\pi \eta W} \over \cosh Wk} \eqno(4.10)$$

\noindent Thus
$$\beta \omega = - \int_{- \infty}^{\infty} {dk \over 2 \pi} \ln \left [ 1 +
\pi \zeta
{e^{\pi \eta W} \over \cosh Wk} \right ] + {1 \over 2} \pi \eta^2 W
\eqno(4.11)$$

For obtaining a large-$W$ expansion of (4.11), we rewrite it as
$$\beta \omega = - \int_0^{\infty} {dk \over \pi} \ln {1 + e^{-2Wk} + 2 \pi
\zeta e^{W(\pi
\eta - k)} \over 1 + e^{-2Wk}} + \int_0^{\pi \eta} {dk \over \pi} \ln e^{W(\pi
\eta - k)}
\eqno(4.12)$$

\noindent extract from the first integral in (4.12) the factor
$$\int_0^{\infty} {dk \over \pi} \ln \left ( 1 + e^{-2Wk} \right ) = {\pi \over
24W}
\eqno(4.13)$$

\noindent and regroup the other terms into two integrals on the $k$-ranges ($0,
\pi \eta $)
and ($\pi \eta , \infty$). After simple changes of variable, neglecting
exponentially small
terms (which allows to extend one of the integration ranges), one obtains
$$\eqalignno{
\beta \omega &\sim - \eta \ln (2 \pi \zeta ) + {\pi \over 24W} - {1 \over \pi
W}
\int_0^{\infty } du \ln \left [ \left (1 + 2 \pi \zeta e^{-u} \right ) \left (
1 + {1 \over 2
\pi \zeta} e^{-u} \right ) \right ] \cr
&= - \eta \ln (2 \pi \zeta ) - {\pi \over 8W} - {1 \over 2 \pi W} \left [ \ln
(2 \pi \zeta )
\right ]^2 &(4.14) \cr
 }$$

\noindent The corresponding particle density is
$$n = - \zeta {d \over d \zeta} (\beta \omega ) \sim \eta + {1 \over \pi W} \ln
(2 \pi \zeta
) \eqno(4.15)$$

\noindent Neutrality still occurs when $2 \pi \zeta = 1$, (3.20) is still
valid, and (4.8b)
is verified by (4.14). \par

The correlation function is still given by (3.21) and (3.23) now with (4.10).
Thus
$$\eqalignno{
g(x, 0) &= \int_{- \infty}^{\infty} {dk \over 2 \pi} {e^{ikx} \over 1 + {1
\over \pi \zeta}
\ e^{-\pi \eta W} \cosh Wk} \sim \int_0^{\infty} {dk \over \pi} {\cos kx \over
1 + 2e^{-\pi
n W} \cosh Wk} \cr
&\sim {\sin (\pi nx) \over W \sinh {\pi x \over W}} &(4.16) \cr
 }$$

\noindent and
$$\beta S(x) = - {1 - \cos (2 \pi nx) \over W^2 \sinh^2 {\pi x \over W}}
\eqno(4.17)$$

\noindent When the oscillations of (4.17) are averaged out, the universal
expression (4.4) is
verified. \par \medskip

\noindent {\bf 5. COULOMB SYSTEM IN A PLANE PARALLEL TO AN IDEAL\break
\noindent CONDUCTOR}
\medskip

{\bf 5.1. \underbar{The system}} \medskip
The universal properties discussed in Section 2 to 4 can be generalized to
systems of higher
dimension. As an example, in three-dimensional space $xyz$, we consider a
conducting
classical system of charges confined in the plane $z = W$ and kept at an
average potential
$\Phi$, while the plane $z = 0$ is an ideal conductor at potential zero. Such a
model might
be relevant for describing electrons trapped at the surface of liquid helium in
front of an
electrode located under the surface.$^{(11)}$ We transpose the derivations and
results of
Section~3 to the present system. \par

We note a position as ${\bf r} = (x, y, z) = ({\bfrho} , z)$ where ${\bfrho} =
(x, y)$. In the
half-space $z > 0$, the Green function $G_0$, solution of
$$\Delta G_0({\bf r}, {\bf r}') = - 4 \pi \delta ({\bf r} - {\bf r}')
\eqno(5.1)$$

\noindent constrained to vanish on the plane $z = 0$ and at infinity is
$$G_0({\bf r}, {\bf r}') = {1 \over |{\bf r} - {\bf r}'|} - {1 \over |{\bf r} -
{\bf r}'^*|}
\eqno(5.2)$$

\noindent where ${\bf r}'^* = (x', y' - z')$ is the image of ${\bf r}'$. In the
plane $z = W$,
the interaction is
$$v({\bfrho} , {\bfrho} ') = {1 \over |{\bfrho} - {\bfrho} '|} - {1 \over
[({\bfrho} -
{\bfrho} ')^2 + 4W^2]^{1/2}} \eqno(5.3)$$ \par \medskip

{\bf 5.2. \underbar{Correlations}} \medskip
The Green function which vanishes on both the $z = 0$ and $z = W$ planes is
$$G({\bf r}, {\bf r}') = \sum_{n = - \infty}^{\infty} G_0 ({\bf r}, {\bf r}' +
2n W {\bf u})
\quad {\rm if} \ 0 < z, z' < W \eqno(5.4a)$$

\noindent where ${\bf u}$ is the unit vector along the $z$ axis,
$$G({\bf r}, {\bf r}') = {1 \over |{\bf r} - {\bf r}'|} - {1 \over \left [
({\bfrho} - {\bfrho}
')^2 + (z + z' - 2W)^2 \right ]^{1/2}} \quad {\rm if} \ z, z' > W \eqno(5.4b)$$

$$G({\bf r}, {\bf r}') = 0 \quad {\rm if} \ 0 < z < W, \ z' > W \ ; \ {\rm or}
\ z > W, \ O <
z' < W \eqno(5.4c)$$

\noindent The potential correlation function is given by (2.7), and from the
analog of (2.8)
one finds for the surface charge $\sigma ({\bf r})$ correlation function in the
plane $z = W$
$$\beta S ({\bfrho} - {\bfrho} ') := \beta < \sigma ({\bfrho} ) \sigma
({\bfrho} ')>^T = - {1
\over 4 \pi^2} \sum_{n=0}^{\infty} {({\bfrho} - {\bfrho} ')^2 - 2(2nW)^2 \over
\left [
({\bfrho} - {\bfrho} ')^2 + (2nW)^2 \right ]^{5/2}} \eqno(5.5)$$

\noindent At short distances, $\beta S \sim - 1/4\pi^2 ({\bfrho} - {\bfrho}
')^3$, in agreement
with the formula (1.1) for a plane alone~; at long distances, the
Euler-MacLaurin summation
formula gives $\beta S \sim - 1/8 \pi^2 ({\bfrho} - {\bfrho} ')^3$. \par

An alternative direct derivation of (5.5) uses (2.9) and the two-dimensional
Fourier transform
$$\tilde{v}(k) = {2 \pi \over |{\bf k}|} \left ( 1 - e^{-2W|{\bf k}|} \right )
\eqno(5.6)$$
\par \medskip

{\bf 5.3. \underbar{Free energy or grand potential}} \medskip
In the region $0 < z < W$, using (2.7) gives for the $T_{zz}$ component of the
Maxwell stress
tensor, the force per unit area,
$$T_{zz} := {1 \over 4 \pi} <E_z^2 - {1 \over 2} {\bf E}^2> = {k_B T \zeta (3)
\over 8 \pi
W^3} + {\Phi^2 \over 8 \pi W^2} \eqno(5.7)$$

\noindent where $\zeta (3)$ = 1.202 ... is a value of the Rieman zeta function.
The last term
of (5.7) is the standard attractive force between the plates of a plane
condenser. The
(usually much smaller) thermal term $k_BT \zeta (3)/8 \pi W^3$ can be obtained
as the
classical limit of the celebrated more general theory$^{(12)}$ of Van der Waals
type forces
between macroscopic bodies.$^4$\footnote{}{$^4$ Our result agrees with eq.
(5.5) of ref. 12,
with $\varepsilon_0 = \infty$ and the integral evaluated exactly.} \par

By integrating $(\partial f/\partial W)_{\sigma} = (\partial \omega /\partial
W)_{\Phi} = T_{yy}$,
one obtains $$\beta f(W) = \beta f(\infty ) - {\zeta (3) \over 16 \pi W^2} +
{\Phi^2 \over 8 \pi
W} \eqno(5.8a)$$
$$\beta \omega (W) = \beta f(\infty ) -{\zeta (3) \over 16 \pi W^2} - {\Phi^2
\over 8 \pi W}
\eqno(5.8b)$$

Alternatively the thermal term of (5.8) can be obtained by using (2.9), the
analog of
(2.14), and (5.6). \par

For the $d$-dimensional analog of the present system, the thermal part of the
force per unit
area is found to be
$${k_BT (d - 1) \Gamma \left ( {d \over 2} \right ) \zeta (d) \over \pi^{d/2}
(2W)^d}
\eqno(5.9)$$ \par \medskip

\noindent {\bf 6. CONCLUSION}
\medskip
The occurrence of universal properties in conducting classical Coulomb systems
is especially
visible in systems of restricted dimensionality. The simplest, already
known,\break \noindent
example is the smoothed charge-charge correlation function $S(r)$ for particles
in a plane with
$1/r$ interactions, as given by eq. (1.1). If (1.1) is understood in the sense
of distributions,
the prescription for regularizing its integral is
$$\int S(r) d^3{\bf r} = - {k_BT \over 4 \pi^2} \int {d^3{\bf r} \over r^3} = 0
\eqno(6.1)$$

\noindent and therefore (1.1) even grossly represents the full $S(r)$,
including its
self-part term, since (6.1) correctly expresses the screening rule. \par

In the present paper, we have derived universal smoothed charge correlation
functions for
other geometries involving some boundary conditions for the electric potential
outside the
system of charges. These boundary conditions involve some macroscopic length
scale $W$, and
we have exhibited universal $W$-dependences of the free energy and the grand
potential. \par

In Sections 2 to 4, we have considered one-dimensional systems with a
two-dimensional Coulomb
interaction, because, for these systems, we had at hand exactly solvable models
on which we
could test our generic results. However, similar generic results can be easily
derived for
any dimension, and in Section~5 we have given an example involving the usual
three-dimensional Coulomb law. \par

A by-product of Section~4 is a new one-dimensional solvable model~: particles
on a line,
interacting through the potential $- e^2 \ln |\tanh [(\pi /2W)(x - x')]|$,
without any
background nor self-energy. When $\Gamma := \beta e^2 = 2$, the correlations
and the
thermodynamics can be obtained exactly, by a minor adaptation of Section~4.

  \vfill\supereject \noindent{\bf REFERENCES}
\medskip \item{1.} P. J. Forrester, {\it J. Stat. Phys.} {\bf 54} : 57 (1989).
\item{2.} B.
Jancovici, {\it J. Stat. Phys.} {\bf 80} : 445 (1995), and references quoted
there.  \item{3.}
P. J. Forrester, {\it J. Stat. Phys.} {\bf 63} : 491 (1991). \item{4.} B.
Jancovici, G.
Manificat, and C. Pisani, {\it J. Stat. Phys.} {\bf 76} : 307 (1994).
\item{5.} B. Jancovici and G. T\'ellez, {\it J. Stat. Phys.}, to be published.
\item{6.} P. J. Forrester, {\it J. Stat. Phys.} {\bf 76} : 331 (1994).
\item{7.} P. J. Forrester, {\it SIAM J. Math. Anal.} {\bf 21} : 270 (1990).
\item{8.} M. Gaudin, {\it Nucl. Phys.} {\bf 85} : 545 (1966).
\item{9.} P.J. Forrester, {\it Phys. Lett. A}  {\bf 173} : 355 (1993).
\item{10.} P. J. Forrester, {\it J. Phys. A} {\bf 18} : 1419 (1985).
\item{11.} F.I.B. Williams, {\it J. Phys. IV} (France) {\bf C2} : 3 (1993), and
references
quoted there.
\item{12.} E. M. Lifshitz, {\it Sov. Phys. JETP} {\bf 2} : 73 (1956).

\bye